\documentclass[prd,preprint,showpacs]{revtex4-1}
\usepackage[utf8]{inputenc}
\usepackage{amsmath}
\usepackage{latexsym}
\usepackage{amsfonts}
\usepackage{graphicx}
\usepackage{mathrsfs}
\usepackage{CJK}
\usepackage{longtable}

\usepackage{hyperref}
\hypersetup{
  colorlinks = true,
  urlcolor = blue,
  linkcolor = blue,
  citecolor = green,
  filecolor = magenta,
}

\newcommand{\ud}{\mathrm{d}}

\begin{abstract}
The direct detection of gravitational waves by LIGO/Virgo opened the possibility to test General Relativity and its alternatives in the high speed, strong field regime.
Alternative theories of gravity generally predict more polarizations than General Relativity, so it is important to study the polarization contents of theories of gravity to reveal the nature of gravity.
In this talk, we analyzed the polarizations contents of Horndeski theory and $f(R)$ gravity.
We found out that in addition to the familiar plus and cross polarizations, a \emph{massless} Horndeski theory predicts an extra transverse polarization, and  there is a mix of the pure longitudinal and transverse breathing polarizations in the \emph{massive} Horndeski theory and $f(R)$ gravity.
It is possible to use pulsar timing arrays to  detect the extra polarizations in these theories.
We also pointed out that the classification of polarizations using Newman-Penrose variables cannot be applied to the massive modes.
It cannot be used to classify polarizations in Einstein-\ae ther theory or generalized TeVeS theory, either.
\end{abstract}

\begin{document}

\title{The Polarizations of Gravitational Waves}
%\author{Yungui Gong(龚云贵)}
\author{Yungui Gong}
\email{yggong@hust.edu.cn}
\thanks{talk given by this author.}
\affiliation{School of Physics, Huazhong University of Science and Technology, Wuhan, Hubei 430074, China}
%\author{Shaoqi Hou(侯绍齐)}
\author{Shaoqi Hou}
\email{shou1397@hust.edu.cn}
\thanks{corresponding author.}
\affiliation{School of Physics, Huazhong University of Science and Technology, Wuhan, Hubei 430074, China}

\maketitle

\section{Introduction}

The detection of gravitational waves (GWs) by LIGO Scientific and Virgo collaborations further supports  General Relativity (GR) and provides a new tool to study gravitational physics in the high speed, strong field regime \cite{Abbott:2016blz,Abbott:2016nmj,Abbott:2017vtc,Abbott:2017oio,TheLIGOScientific:2017qsa,Abbott:2017gyy}.
In order to confirm GWs predicted by GR, it is necessary to determine the polarizations of GWs.
In GR, the GW has two polarization states, the plus and cross modes.
In contrast, alternative metric theories of gravity predict up to six polarizations \cite{Will:2014kxa}, so
the detection of the polarizations of GWs can be used to probe the nature of gravity \cite{Isi:2015cva,Isi:2017equ}.
This can be done by the network of  Advanced LIGO (aLIGO) and Virgo,
LISA \cite{Audley:2017drz} and TianQin \cite{Luo:2015ght}, and
pulsar timing arrays \cite{Hobbs:2009yy,Kramer:2013kea}, etc..
In fact,  GW170814 was the first GW event to test the polarization content of GWs.
The analysis revealed that the pure tensor polarizations were favored against pure vector and pure scalar polarizations \cite{Abbott:2017oio,Abbott:2017tlp}.
With the advent of more advanced detectors, there exists a better chance to pin down the polarization content and thus, the nature of gravity in the future.

The six polarizations of the null plane GWs are classified by the little group $E(2)$
of the Lorentz group with the help of the six independent Newman-Penrose (NP) variables $\Psi_2$, $\Psi_3$, $\Psi_4$ and $\Phi_{22}$ \cite{Newman:1961qr,Eardley:1974nw,Eardley:1973br}.
In particular, the complex variable $\Psi_4$ denotes the  plus and cross polarizations,
$\Phi_{22}$ represents the transverse breathing polarization, the complex variable $\Psi_3$ corresponds to the vector-x and vector-y polarizations,
and $\Psi_2$ corresponds to the longitudinal polarization.
%Under the $E(2)$ transformation, all other modes can be generated from $\Psi_2$, so if $\Psi_2\neq 0$, then we may see all six modes in some coordinates.
%The $E(2)$ classification tells us the general polarization states, but it fails to tell us the correspondence between the polarizations and gravitational theories.
For example, in Brans-Dicke theory \cite{Brans:1961sx}, in addition to the plus and cross modes $\Psi_4$ of the massless gravitons, there exists another breathing mode $\Phi_{22}$ due to the massless Brans-Dicke scalar field \cite{Eardley:1974nw}.

Horndeski theory is the most general scalar-tensor theory of gravity whose action has higher derivatives of the metric tensor $g_{\mu\nu}$ and a scalar field $\phi$, but the equations of motion are at most the second order \cite{Horndeski:1974wa}.
So there is no Ostrogradsky instability \cite{Ostrogradsky:1850fid},
and there are three physical degrees of freedom (d.o.f.).
It is expected that there is an extra polarization state.
If the scalar field is massless, then the additional polarization state should be the breathing mode $\Phi_{22}$ as in Brans-Dicke theory.

The general nonlinear $f(R)$ gravity \cite{Buchdahl:1983zz} is equivalent to a scalar-tensor theory of gravity \cite{OHanlon:1972xqa,Teyssandier:1983zz}.
The equivalent scalar field is  massive, and it excites both the longitudinal and transverse breathing modes \cite{Corda:2007hi,Corda:2007nr,Capozziello:2008rq}.
The analysis shows that there are the plus and the cross polarization, and the polarization state of the equivalent massive field is the mixture of the longitudinal and the transverse breathing polarizations \cite{Liang:2017ahj}.
%Furthermore, the longitudinal polarization is independent of the effective mass of the equivalent scalar field,
%so it cannot be used to understand how the polarization reduces to the transverse breathing mode in the massless limit.

We will show that the classification based on $E(2)$ symmetry cannot be applied to the \emph{massive} Horndeski and $f(R)$ gravity, as there are massive modes in these two theories.
In fact, it cannot be used to classify the polarizations in Einstein-\ae ther theory \cite{Jacobson:2004ts} or generalized TeVeS  theory \cite{Bekenstein:2004ne,Seifert:2007fr,Sagi:2010ei}, as the local Lorentz invariance is violated in both theories \cite{Gong:2018cgj}.

The talk is organized as follows.
Section \ref{sec-e2} briefly reviews the $E(2)$ classification for classifying the polarizations of null GWs.
In Section \ref{sec-fr}, the GW polarization content of $f(R)$ gravity is obtained.
In Section \ref{sec-horn}, the polarization content of Horndeski theory is discussed.
Section \ref{sec-eae-gtvs} discusses the polarization contents of Einstein-\ae ther theory and the generalized TeVeS theory.
Finally,  Section \ref{sec-con} is a brief summary.
In this talk, we use the natural units and the speed of light in vacuum $c=1$.

\section{Review of $E(2)$ Classification}\label{sec-e2}

$E(2)$ classification is a model-independent framework \cite{Eardley:1974nw,Eardley:1973br} to classify the null GWs in a generic, local Lorentz invariant metric theory of gravity using the Newman-Penrose formalism \cite{Newman:1961qr}.
The quasiorthonormal, null tetrad basis $E^\mu_a=(k^\mu,l^\mu,$ $m^\mu,\bar{m}^\mu)$ is chosen to be
\begin{equation}\label{nullb}
  k^\mu=\frac{1}{\sqrt{2}}(1,0,0,1), \,
  l^\mu=\frac{1}{\sqrt{2}}(1,0,0,-1), \,
  m^\mu=\frac{1}{\sqrt{2}}(0,1,i,0), \,
  \bar m^\mu=\frac{1}{\sqrt{2}}(0,1,-i,0),
\end{equation}
where bar  means the complex conjugation.
They satisfy $-k^\mu l_\mu=m^\mu\bar m_\mu=1$ and all other inner products are zero.
Since the null GW propagates in the $+z$ direction, the Riemann tensor is a function of the retarded time $u=t-z$, which implies that $R_{abcd,p}=0$, where $(a,b,c,d)$ range over $(k,l,m,\bar m)$ and $(p,q,r,\cdots)$ range over ($k,m,\bar m$).
The linearized Bianchi identity and the symmetry properties of $R_{abcd}$ imply that
there are six independent nonzero components and they can be written in terms of the following  NP variables,
\begin{equation}\label{eq-innp}
    \Psi_2=-\frac{1}{6}R_{klkl},\,
    \Psi_3=-\frac{1}{2}R_{kl\bar ml},\,
    \Psi_4=-R_{\bar ml\bar ml},\,
    \Phi_{22}=-R_{ml\bar ml},
\end{equation}
and the remaining nonzero NP variables are $\Phi_{11}=3\Psi_2/2$, $\Phi_{12}=\bar\Phi_{21}=\bar\Psi_3$ and $\Lambda=\Psi_2/2$.
Note that $\Psi_2$ and $\Phi_{22}$ are real while $\Psi_3$ and $\Psi_4$ are complex.

These four NP variables $\{\Psi_2,\Psi_3,\Psi_4,\Phi_{22}\}$ can be classified according to their transformation properties under the group $E(2)$.
Under $E(2)$ transformation,
\begin{gather}\label{eq-e2tnp}
    \Psi'_2=\Psi_2,\,
    \Psi'_3=e^{-i\vartheta}(\Psi_3+3\bar\rho\Psi_2),\\
    \Psi'_4=e^{-i2\vartheta}(\Psi_4+4\bar\rho\Psi_3+6\bar\rho^2\Psi_2),\,
    \Phi'_{22}=\Phi_{22}+2\rho\Psi_3+2\bar\rho\bar\Psi_3+6\rho\bar\rho\Psi_2,
\end{gather}
where $\vartheta\in[0,2\pi)$ parameterizes a rotation around the $+z$ direction and the complex number $\rho$ parameterizes a translation in the Euclidean 2-plane.
Based on this, six classes are defined as follows \cite{Eardley:1974nw}.
\begin{description}
  \item[Class II$_6$] $\Psi_2\ne0$. All observers measure the same nonzero amplitude of the $\Psi_2$ mode,
  but the presence or absence of all other modes is observer-dependent.
  \item[Class III$_5$] $\Psi_2=0$, $\Psi_3\ne0$. All observers measure the absence of the $\Psi_2$ mode and the presence of the $\Psi_3$ mode, but
  the presence or absence of $\Psi_4$ and $\Phi_{22}$ is observer-dependent.
  \item[Class N$_3$] $\Psi_2=\Psi_3=0,\,\Psi_4\ne0\ne\Phi_{22}$. The presence or absence of all modes is observer-independent.
  \item[Class N$_2$] $\Psi_2=\Psi_3=\Phi_{22}=0,\,\Psi_4\ne0$. The presence or absence of all modes is observer-independent.
  \item[Class O$_1$] $\Psi_2=\Psi_3=\Psi_4=0,\,\Phi_{22}\ne0$. The presence or absence of all modes is observer-independent.
  \item[Class O$_0$] $\Psi_2=\Psi_3=\Psi_4=\Phi_{22}=0$. No wave is observed.
\end{description}
Note that by setting $\rho=0$ in Eq.~\eqref{eq-e2tnp}, one finds out that $\Psi_2$ and $\Phi_{22}$ have helicity 0, $\Psi_3$ has helicity 1 and $\Psi_4$ has helicity 2.

The relation between $\{\Psi_2,\Psi_3,\Psi_4,\Phi_{22}\}$ and the polarizations of the GW can be found by examining the linearized geodesic deviation equation in the Cartesian coordinates \cite{Eardley:1974nw},
\begin{equation}\label{geodevl}
  \ddot x^j=\frac{d^2x^j}{dt^2}=-R_{tjtk}x^k,
\end{equation}
where $x^j$ represents the deviation vector between two nearby geodesics and $j,\,k=1,2,3$.
The so-called electric component $R_{tjtk}$  is given by the following matrix,
\begin{equation}\label{eq-rtjtkm}
  R_{tjtk}=\left(
  \begin{array}{ccc}
    -\frac{1}{2}(\Re\Psi_4+\Phi_{22}) & \frac{1}{2}\Im\Psi_4 & -2\Re\Psi_3 \\
    \frac{1}{2}\Im\Psi_4 & \frac{1}{2}(\Re\Psi_4-\Phi_{22}) & 2\Im\Psi_3 \\
    -2\Re\Psi_3 & 2\Im\Psi_3 & -6\Psi_2
  \end{array}
  \right),
\end{equation}
where $\Re$ and $\Im$ stand for the real and imaginary parts.
Therefore, $\Re\Psi_4$ and $\Im\Psi_4$ represent the plus and the cross polarizations, respectively;
$\Phi_{22}$ donates the transverse breathing polarization, and $\Psi_2$ donates the longitudinal polarization; finally,
$\Re\Psi_3$ and $\Im\Psi_3$ stand for vector-$x$ and vector-$y$ polarizations, respectively.
In terms of $R_{tjtk}$, the plus mode is represented by $\hat{P}_+=-R_{txtx}+R_{tyty}$, the cross mode is represented by $\hat{P}_\times=R_{txty}$, the transverse breathing mode is donated by $\hat{P}_b=R_{txtx}+R_{tyty}$, the vector-$x$ mode is donated by $\hat{P}_{xz}=R_{txtz}$,
the vector-$y$ mode is given by $\hat{P}_{yz}=R_{tytz}$,
and the longitudinal mode is given by $\hat{P}_l=R_{tztz}$.
For null GWs, the four NP variables $\{\Psi_2,\Psi_3,\Psi_4,\Phi_{22}\}$ with six
independent components are related with the six electric components of Riemann tensor,
and they provide the six independent polarizations $\{\hat{P}_+,\hat{P}_\times,\hat{P}_b,\hat{P}_{xz},\hat{P}_{yz},\hat{P}_l\}$.
By $E(2)$ classification, the longitudinal mode with a nonzero $\Psi_2$ belongs to the most general class $\text{II}_6$.
The presence of the longitudinal mode means that all six polarizations are present in some coordinate systems.

One can apply this framework to discuss some specific metric theories of gravity.
For Brans-Dicke theory, one gets
\begin{equation}\label{eq-rtjtkm-bd}
  R_{tjtk}^\mathrm{BD}=\left(
  \begin{array}{ccc}
    -\frac{1}{2}(\Re\Psi_4+\Phi_{22}) & \frac{1}{2}\Im\Psi_4 & 0 \\
    \frac{1}{2}\Im\Psi_4 & \frac{1}{2}(\Re\Psi_4-\Phi_{22}) & 0 \\
    0 & 0 & 0
  \end{array}
  \right).
\end{equation}

In the next sections, the plane GW solutions to the linearized equations of motion will be obtained for $f(R)$ gravity, Horndeski theory, Einstein-\ae{}ther theory and generalized TeVeS theor.
Then the polarization contents will be determined.
It will show that $E(2)$ classification cannot be applied to the massive mode in $f(R)$ gravity or Horndeski theory.
It cannot be applied to the local Lorentz violating theories, such as Einstein-\ae{}ther theory and generalized TeVeS theory, either.

\section{Gravitational Wave Polarizations in $f(R)$ Gravity}\label{sec-fr}

The action of $f(R)$ gravity is \cite{Buchdahl:1983zz},
\begin{equation}\label{fract}
  S=\frac{1}{2\kappa}\int d^4x\sqrt{-g}f(R).
\end{equation}
It is equivalent to a scalar-tensor theory, since the action can be reexpressed as \cite{OHanlon:1972xqa,Teyssandier:1983zz}
\begin{equation}\label{frst}
  S=\frac{1}{2\kappa}\int d^4x\sqrt{-g}[f(\varphi)+(R-\varphi)f'(\varphi)],
\end{equation}
where $f'(\varphi)=d f(\varphi)/d\varphi$.
The variational principle leads to the following equations of motion,
\begin{equation}
\label{freinseq1}
f'(R)R_{\mu\nu}-\frac{1}{2}f(R)g_{\mu\nu}-\nabla_\mu\nabla_\nu f'(R)+g_{\mu\nu}\Box f'(R)=0,
\end{equation}
where $\Box=g^{\mu\nu}\nabla_\mu\nabla_\nu$.
Taking the trace of Eq. \eqref{freinseq1}, one obtains
\begin{equation}
\label{freinseq2}
f'(R)R+3\Box f'(R)-2f(R)=0.
\end{equation}
For the particular model $f(R)=R+\alpha R^2$, Eq. \eqref{freinseq1} becomes
\begin{equation}
\label{frperteq3}
R_{\mu\nu}-\frac{1}{2}\eta_{\mu\nu}R-2\alpha\left(\partial_\mu\partial_\nu R-\eta_{\mu\nu}\Box R\right)=0,
\end{equation}
Taking the trace of Eq. \eqref{frperteq3} or using Eq. \eqref{freinseq2}, one gets
\begin{equation}
\label{frperteq4}
(\Box-m^2)R=0,
\end{equation}
where $m^2=1/(6\alpha)$ with $\alpha>0$. The graviton mass $m$ has been bounded from above by GW170104 as
$m<m_b=7.7\times10^{-23}\text{ eV}/c^2$ \cite{Abbott:2017vtc}, and the observation of the dynamics of the galaxy cluster puts a more stringent limit, $m<2\times10^{-29}\text{ eV}/c^2$ \cite{Goldhaber1974}.

Now, we want to obtain the GW solutions in the flat spacetime background, so we perturb the metric around the Minkowski metric $g_{\mu\nu}=\eta_{\mu\nu}+h_{\mu\nu}$ to the first order of $h_{\mu\nu}$, and introduce an auxiliary metric tensor
\begin{equation}
\label{canhtt}
\bar{h}_{\mu\nu}=h_{\mu\nu}-\frac{1}{2}\eta_{\mu\nu}h-2\alpha\eta_{\mu\nu} R.
\end{equation}
In an infinitesimal coordinate transformation $x^\mu\rightarrow x'^\mu=x^\mu+\epsilon^\mu$, this tensor transforms according to
\begin{equation}
\label{coortranfeq23}
\bar{h}_{\mu\nu}'=\bar{h}_{\mu\nu}-\partial_\mu\epsilon_\nu-\partial_\nu\epsilon_\mu+\eta_{\mu\nu}\partial_\rho\epsilon^\rho.
\end{equation}
So choose the transverse traceless gauge condition
\begin{equation}
\label{gaugeeq1}
\partial^\mu \bar{h}_{\mu\nu}=0,\quad \bar{h}=\eta^{\mu\nu}\bar{h}_{\mu\nu}=0.
\end{equation}
In this gauge, one obtains
\begin{equation}
\label{frperteq8}
\Box\bar{h}_{\mu\nu}=0.
\end{equation}
Therefore, the equations of motion are Eqs. (\ref{frperteq4}) and (\ref{frperteq8}).

The plane wave solution are given below,
\begin{gather}
  \bar h_{\mu\nu}=e_{\mu\nu}\exp(iq_\mu x^\mu)+c.c.,\label{sols-h}\\
  R=\phi_1\exp(ip_\mu x^\mu)+c.c.,\label{sols-r}
\end{gather}
where $c.c.$ stands for the complex conjugation, $e_{\mu\nu}$ and $\phi_1$ are the amplitudes with $q^\nu e_{\mu\nu}=0$ and $\eta^{\mu\nu}e_{\mu\nu}=0$, and $q_\mu$ and $p_\mu$ are the wave numbers satisfying
\begin{equation}\label{wvnm}
  \eta^{\mu\nu}q_\mu q_\nu=0, \quad
  \eta^{\mu\nu}p_\mu p_\nu=-m^2.
\end{equation}

\subsection{Physical Degrees of Freedom}

In this subsection, we will find the number of physical degrees of freedom in $f(R)$ gravity,
using  the Hamiltonian analysis.
It is convenient to carry out the Hamiltonian analysis with the action (\ref{frst}).
With the Arnowitt-Deser-Misner (ADM) foliation \cite{Arnowitt:1962hi}, the metric takes the following form
\begin{equation}\label{lem}
  ds^2=-N^2dt^2+h_{jk}(dx^j+N^jdt)(dx^k+N^kdt),
\end{equation}
where $N,N^j,h_{jk}$ are the lapse function, the shift function and the induced metric on the constant $t$ slice $\Sigma_t$, respectively. Let $n_\mu=-N\nabla_\mu t$ be the unit normal to $\Sigma_t$, and  $K_{\mu\nu}=\nabla_\mu n_\nu+n_\mu n^\rho\nabla_\rho n_\nu$ is the exterior curvature. In terms of ADM variables and setting $\kappa=1$ for simplicity, the action (\ref{frst}) becomes
\begin{equation}
  S = \int d^4xN\sqrt{h}\Big[\frac{1}{2}f'(\mathscr R-\varphi)+\frac{1}{2}f+\frac{1}{2}f'(K_{jl}K^{jl}-K^2)+\frac{K}{N}(N_jD^jf'-f''\dot\varphi)+D_j f' D^j\ln N\Big],
\end{equation}
where  $\mathscr R$ is the Ricci scalar for $h_{jk}$ and $K=h^{jk}K_{jk}$. In this action, there are totally 11 dynamical variables: $N,N_j,h_{jk}$ and $\varphi$.
Four primary constraints are,
\begin{equation}\label{prcon}
    \pi^N = \frac{\delta S}{\delta \dot N}\approx0,\quad  \pi^j=\frac{\delta S}{\delta \dot N_j}\approx0.
\end{equation}
The conjugate momenta for $h_{jk}$ and $\varphi$ can also be calculated, and the Legendre transformation leads to the following Hamiltonian,
\begin{equation}
  H=\int_{\Sigma_t}d^3x\sqrt{h}(NC+N_jC^j),
\end{equation}
where the boundary terms have been ignored.
Then, the consistence conditions result in four secondary constraints, i.e., $C\approx0$ and $C^j\approx0$, and it can be shown that there are no further secondary constraints.
All the constraints are of the first class, so the number of physical degrees of freedom of $f(R)$ gravity is
\begin{equation}\label{dof}
  n=\frac{22-8\times2}{2}=3,
\end{equation}
as expected.

\subsection{Polarization Content}

To obtain the polarization content of GWs in $f(R)$ gravity, we calculate the geodesic deviation equations. Assume  the GW propagates in the $+z$ direction with the wave vectors given by
\begin{equation}\label{wvecs}
  q^\mu=\omega(1,0,0,1),\quad p^\mu=(\Omega,0,0,\sqrt{\Omega^2-m^2}).
\end{equation}
Inverting Eq.~(\ref{canhtt}), one obtains the metric perturbation,
\begin{equation}\label{metper}
  h_{\mu\nu}=\bar h_{\mu\nu}(t-z)-2\alpha\eta_{\mu\nu}R(vt-z),
\end{equation}
where $v=\sqrt{\Omega^2-m^2}/\Omega$. As expected, $\bar h_{\mu\nu}$ induces the $+$ and $\times$ polarizations.
Now to investigate the polarization state caused by the massive scalar field, set $\bar h_{\mu\nu}=0$.
The geodesic deviation equations are
\begin{equation}\label{geodev}
  \ddot{x}=\alpha\ddot R x,\quad \ddot y=\alpha\ddot{R}y,\quad \ddot z=-\alpha m^2R z=-\frac{1}{6}R z.
\end{equation}
Therefore, the massive scalar field induces a mix of the pure longitudinal and the breathing modes.

The NP formalism \cite{Eardley:1974nw,Eardley:1973br} is not suitable to determine the polarization content
of $f(R)$ gravity because the NP formalism was formulated for null GWs.
Indeed, the calculation shows that $\Psi_2$ is zero, which  means the absence of the longitudinal polarization according to the NP formalism.
However, from Eq. \eqref{geodev} implies the existence of the longitudinal polarization.
Nevertheless, the six polarization states are completely determined by the electric part of the Riemann tensor $R_{tjtk}$, so one can  use the six polarizations classified by the NP formalism as the base states.
In terms of these polarization base states, the massive
scalar field excites a mix of the longitudinal and the breathing modes.
Since one cannot take the massless limit of $f(R)$ gravity, so we consider more general massive scalar-tensor theory of gravity -- Horndeski theory.

\section{Gravitational Wave Polarizations in Horndeski Theory}\label{sec-horn}

The action is given by \cite{Horndeski:1974wa},
\begin{equation}
\label{acth}
  S=\int d^4x\sqrt{-g}(L_2+L_3+L_4+L_5),
\end{equation}
where
\begin{gather*}
L_2=K(\phi,X),\quad L_3=-G_3(\phi,X)\Box \phi, \quad L_4=G_4(\phi,X)R+G_{4,X}\left[(\Box\phi)^2-(\nabla_\mu\nabla_\nu\phi)(\nabla^\mu\nabla^\nu\phi)\right], \nonumber\\
L_5=G_5(\phi,X)G_{\mu\nu}\nabla^\mu\nabla^\nu\phi
-\frac{1}{6}G_{5,X}\left[(\Box\phi)^3-3(\Box\phi)(\nabla_\mu\nabla_\nu\phi)(\nabla^\mu\nabla^\nu\phi)\right.\nonumber\\
\left.+2(\nabla^\mu\nabla_\alpha\phi)(\nabla^\alpha\nabla_\beta\phi)(\nabla^\beta\nabla_\mu\phi)\right].
\end{gather*}
Here, $X=-\nabla_\mu\phi\nabla^\mu\phi/2$, $\Box\phi=\nabla_\mu\nabla^\mu\phi$,
the functions $K$, $G_3$, $G_4$ and $G_5$ are arbitrary functions of $\phi$ and $X$, and $G_{j,X}(\phi,X)=\partial G_j(\phi,X)/\partial X$ with $j=4,5$.
Horndeski theory includes several theories as its subclasses.
For example, one can set $G_3=G_5=0$, $K=f(\phi)-\phi f'(\phi)$ and $G_4=f'(\phi)$ with $f'(\phi)=d f(\phi)/d\phi$ to reproduce $f(R)$ gravity.

\subsection{Gravitational Wave Solutions}

To find the GW solutions in the flat spacetime background, one perturbs the metric tensor and the scalar field such that  $g_{\mu\nu}=\eta_{\mu\nu}+h_{\mu\nu}$ and $\phi=\phi_0+\varphi$ with $\phi_0$ a constant.
The consistence of the equations of motion requires that $K(\phi_0,0)=0$ and $K_{,\phi_0}=\partial K(\phi,X)/\partial\phi|_{\phi=\phi_0,X=0}=0$.
The linearied equations of motion are
\begin{equation}\label{eqht}
(\Box-m^2)\varphi = 0,\quad
G_{\mu\nu}^{(1)}-\frac{G_{4,\phi_0}}{G_4(0)}(\partial_\mu\partial_\nu\varphi-\eta_{\mu\nu}\Box\varphi)=0,
\end{equation}
where $G_4(0)=G_4(\phi_0,0)$, $K_{,X_0}=\partial K(\phi,X)/\partial X|_{\phi=\phi_0,X=0}$ and the mass squared of the scalar field can be easily read off,
\begin{equation}
\label{msq}
m^2=-\frac{K_{,\phi_0\phi_0}}{K_{,X_0}-2G_{3,\phi_0}+3G_{4,\phi_0}^2/G_{4}(0)}.
\end{equation}

Analogously to Eq. (\ref{canhtt}), introduce a field $\tilde h_{\mu\nu}$,
\begin{equation}
\label{auht}
\tilde h_{\mu\nu}=h_{\mu\nu}-\frac{1}{2}\eta_{\mu\nu}\eta^{\alpha\beta}h_{\alpha\beta}-\frac{G_{4,\phi_0}}{G_{4}(0)}\eta_{\mu\nu}\varphi,
\end{equation}
and choose the transverse traceless gauge $\partial_\mu \tilde h^{\mu\nu}=0$, $\eta^{\mu\nu}\tilde h_{\mu\nu}=0$ by
using the gauge freedom, then Eqs.~\eqref{eqht} become two wave equations,
\begin{equation}\label{eq-wave}
(\Box-m^2)\varphi = 0,\,
\Box\tilde h_{\mu\nu} = 0.
\end{equation}

\subsection{Polarization Content}

The similarity between Eq.~\eqref{eq-wave} and Eqs.  \eqref{frperteq4} and \eqref{frperteq8} makes it clear that there are the plus and the cross polarizations, and the massive scalar field $\phi$ excites a mix of the transverse breathing and the longitudinal polarizations.
Let us calculate the electric component $R_{tjtk}$ of the Riemann tensor given below
\begin{equation}\label{eq-rtjtk}
  R_{tjtk}=
  \left(
  \begin{array}{ccc}
    -\frac{1}{2}q_t^2\sigma\varphi+\frac{1}{2}\Omega^2\tilde h_{xx} & \frac{1}{2}\Omega^2\tilde h_{xy} & 0 \\
    \frac{1}{2}\Omega^2\tilde h_{xy} & -\frac{1}{2}q_t^2\sigma\varphi-\frac{1}{2}\Omega^2\tilde h_{xx} & 0 \\
    0 & 0 & -\frac{1}{2}m^2\sigma\varphi
  \end{array}\right).
\end{equation}
for a GW propagating in the $+z$ direction with waves vectors given by Eq. (\ref{wvecs}).
Here, $\sigma=G_{4,\phi_0}/G_4(0)$.
It is clear that $\tilde h_{\mu\nu}$ excites the plus and the cross polarizations by switching off the scalar perturbation $\varphi=0$.
By setting $\tilde h_{\mu\nu}=0$, one can study the polarizations caused by the scalar field.
If the scalar field is massless ($m=0$),
then $R_{tztz}=0$, so the scalar field excites only the transverse breathing polarization ($R_{txtx}=R_{tyty}$).
If $m\ne0$, a Lorentz boost can be performed such that $q_z=0$.
In this frame,
the geodesic deviation equations are,
\begin{equation}\label{geodevsl}
  \ddot x^j=\frac{1}{2}m^2\sigma\varphi x^j,\, j=1,2,3.
\end{equation}
If the initial deviation vector between two geodesics is $x^j_{0}=(x_0,y_0,z_0)$,  integrating the above equations twice to find,
\begin{equation}\label{chcos}
  \delta x^j\approx -\frac{1}{2}\sigma \varphi x^j_{0}.
\end{equation}
Eq.~\eqref{chcos} means that a sphere of test particles would oscillate isotropically in all directions,
so the massive scalar field excites the longitudinal polarization in addition to the breathing polarization.
Note that in the rest frame of a massive field, one cannot take the massless limit, as there is no rest frame for a massless field propagating at the speed of light.
However, the massless limit can be taken in Eq. \eqref{eq-rtjtk} if the rest frame condition $q_t=m$ is not imposed ahead of time.

However, in the actual observation, it is almost impossible for the test particles, such the mirrors in aLIGO/Virgo, to be in the rest frame of the (massive) scalar gravitational wave.
So one should also study how the scalar GW deviates the nearby geodesics in this frame.
In this case, the deviation vector is given by
\begin{equation}
  \delta x\approx -\frac{1}{2}\sigma \varphi x_0,\quad
  \delta y \approx -\frac{1}{2}\sigma \varphi y_0,\quad
  \delta z\approx -\frac{1}{2}\frac{m^2}{q_t^2}\sigma\varphi z_0.\label{chcos-z}
\end{equation}
From this, one clearly finds out that when $m\ne0$, the scalar field excites a mix of the longitudinal and the transverse breathing polarizations, while when $m=0$, it excites only the transverse breathing polarization.

The NP variables can be calculated.
One obtains
\begin{equation}
\label{psi20}
\Psi_2=\frac{1}{12}(R_{txtx}+R_{tyty}-2R_{tztz}
+2R_{xyxy}-R_{xzxz}-R_{yzyz})+\frac{1}{2}iR_{tzxy}
=0,
\end{equation}
and several nonvanishing NP variables
\begin{gather}\label{nps}
 \Psi_4=-\omega^2(\tilde h_{xx}-i\tilde h_{xy}),\quad
\Phi_{22}=\frac{(\Omega+\sqrt{\Omega^2-m^2})^2}{4}\sigma\varphi,
  \\
\Phi_{00}=\frac{4(\Omega-\sqrt{\Omega^2-m^2})^2}{(\Omega+\sqrt{\Omega^2-m^2})^2}\Phi_{22},\quad
\Phi_{11}=-\Lambda=\frac{4m^2}{(\Omega+\sqrt{\Omega^2-m^2})^2}\Phi_{22}.
\end{gather}
Note that  for null gravitational waves only $\Psi_2=-R_{tztz}/6$, and in general case we should use Eq. \eqref{psi20}.
Next, express $R_{tjtk}$  in terms of NP variables as a matrix displayed below,
\begin{equation}\label{rtjtknp}
  R_{tjtk}= \left(
  \begin{array}{ccc}
    \Upsilon-\frac{1}{2}\Re\Psi_4 & \frac{1}{2}\Im\Psi_4 & 0 \\
    \frac{1}{2}\Im\Psi_4 & \Upsilon+\frac{1}{2}\Re\Psi_4 & 0 \\
    0 & 0 & -2(\Lambda+\Phi_{11})
  \end{array}\right),
\end{equation}
with $\Upsilon=-2\Lambda-\frac{\Phi_{00}+\Phi_{22}}{2}$.
The difference from Eq.~\eqref{eq-rtjtkm-bd} shows the failure of NP formalism in classifying the polarizations of the massive mode.

\subsection{Experimental Tests}

\begin{figure}[htp]
  \centering
  \includegraphics[width=0.4\textwidth,clip]{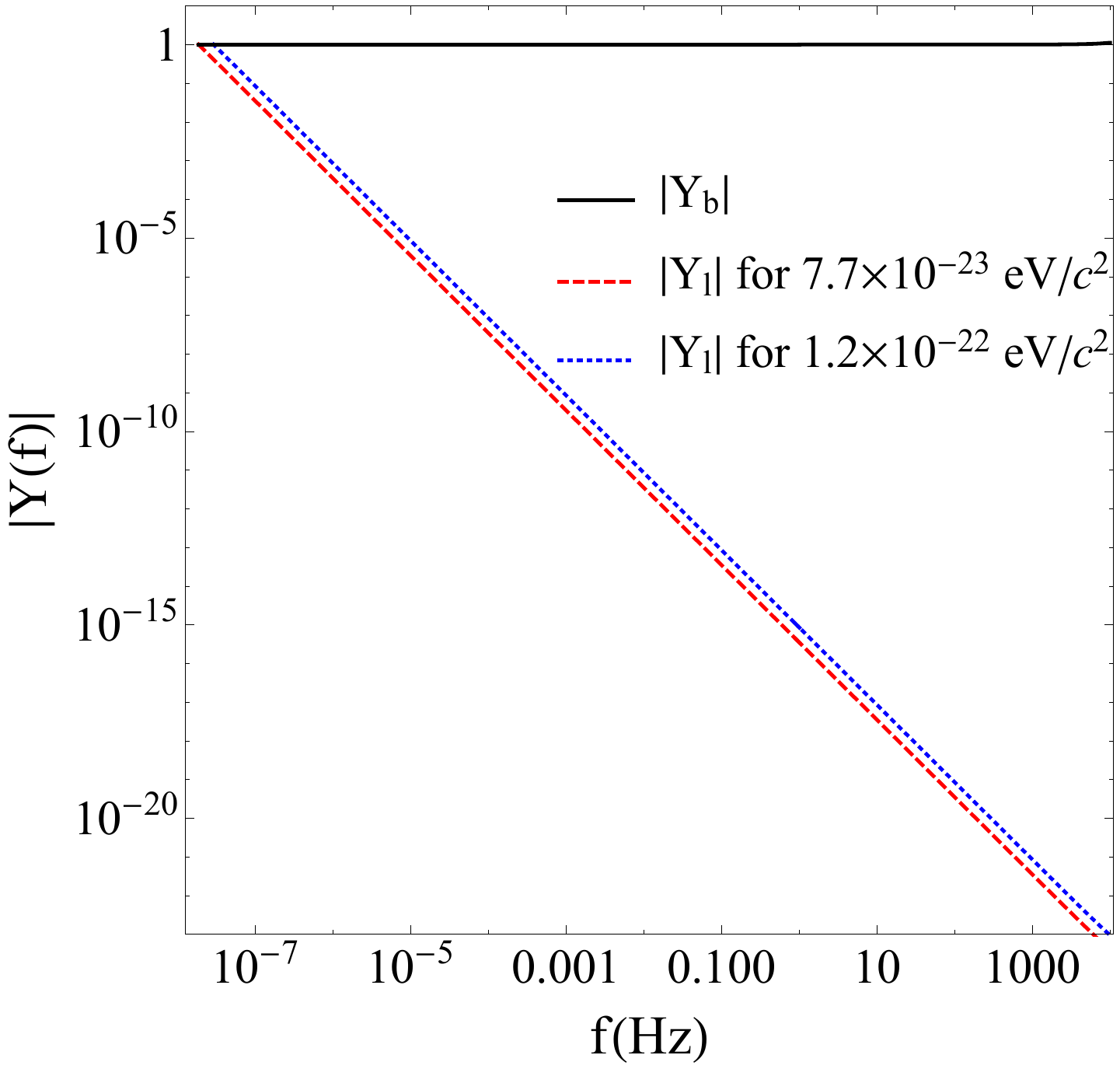}
  \caption{The absolute values of the longitudinal and transverse response functions $|Y_l(f)|$ and $|Y_b(f)|$
  as functions of $f$ for aLIGO  to a scalar gravitational wave with the masses $1.2\times10^{-22}\,\text{eV}/c^2$ \cite{Abbott:2016blz} (dotted blue curve) and $7.7\times 10^{-23}$ eV/$c^2$ \cite{Abbott:2017vtc} (dashed red curve).  The solid curve denotes $|Y_b(f)|$.
  Taken from Ref.~\cite{Gong:2017bru}.}\label{fig-yl}
\end{figure}

The interferometers can detect GWs by measuring the change in the propagation time of photons.
The interferometer response function is important \cite{Rakhmanov:2004eh,Corda:2007hi}.
Figure \ref{fig-yl} shows the absolute value of the longitudinal and transverse response functions for
aLIGO  to a scalar GW with the masses $1.2\times10^{-22}\,\text{eV}/c^2$ \cite{Abbott:2016blz} and $7.7\times 10^{-23}$ eV/$c^2$ \cite{Abbott:2017vtc}.
This graph shows it is difficult to test the existence of longitudinal polarization by interferometer such as aLIGO.

\begin{figure}
  \centering
  \includegraphics[width=0.4\textwidth]{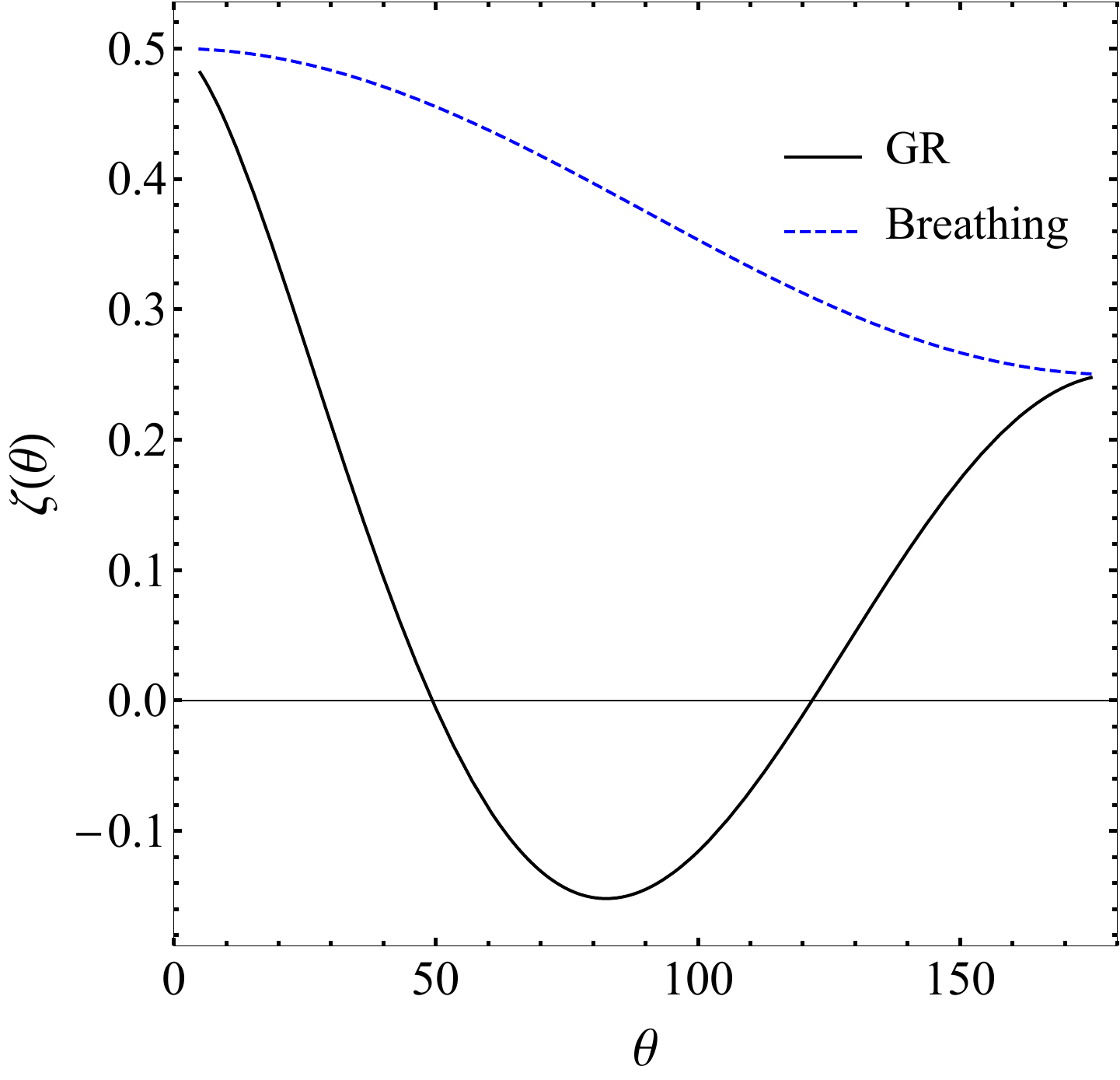}
  \includegraphics[width=0.4\textwidth]{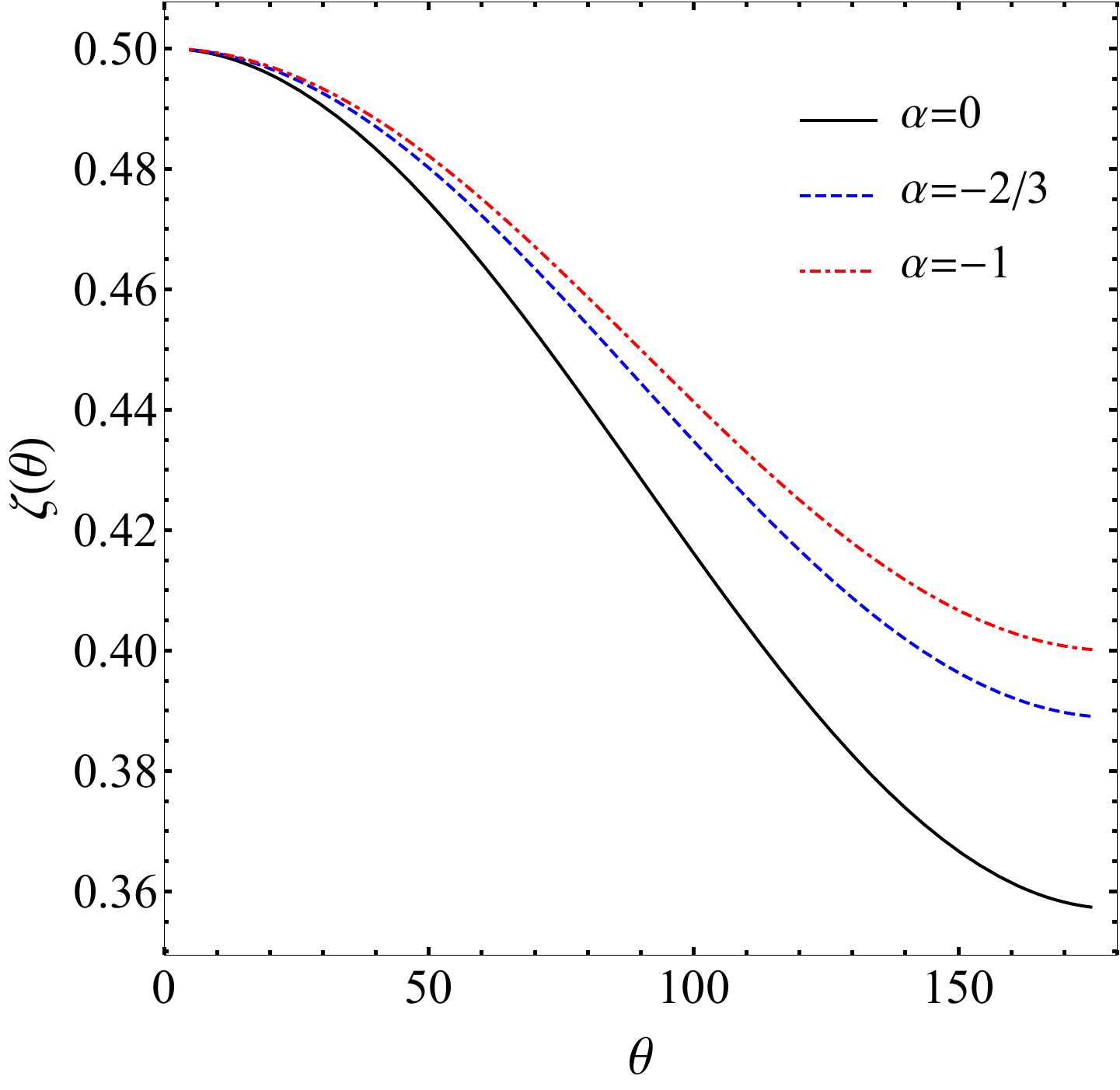}
  \caption{The normalized cross-correlation functions $\zeta(\theta)=C(\theta)/C(0)$.
  The left panel shows the cross-correlations when the scalar field is massless, i.e.,
  when there is no longitudinal polarization.
  The solid curve is for  the plus or the cross polarizations,
  and the dashed curve for the transverse breathing polarization.
  The right panel shows the cross-correlations induced together by the transverse breathing and longitudinal polarizations when the mass
  of the scalar field is taken to be $m_b=7.7\times10^{-23}\,\mathrm{eV}/c^2$.
  The calculation was done assuming the observation time $T=5$ yrs.
  Taken from Ref.~\cite{Hou:2017bqj}.}\label{fig-normcorr}
\end{figure}

\begin{figure}
  \centering
  \includegraphics[width=0.45\textwidth]{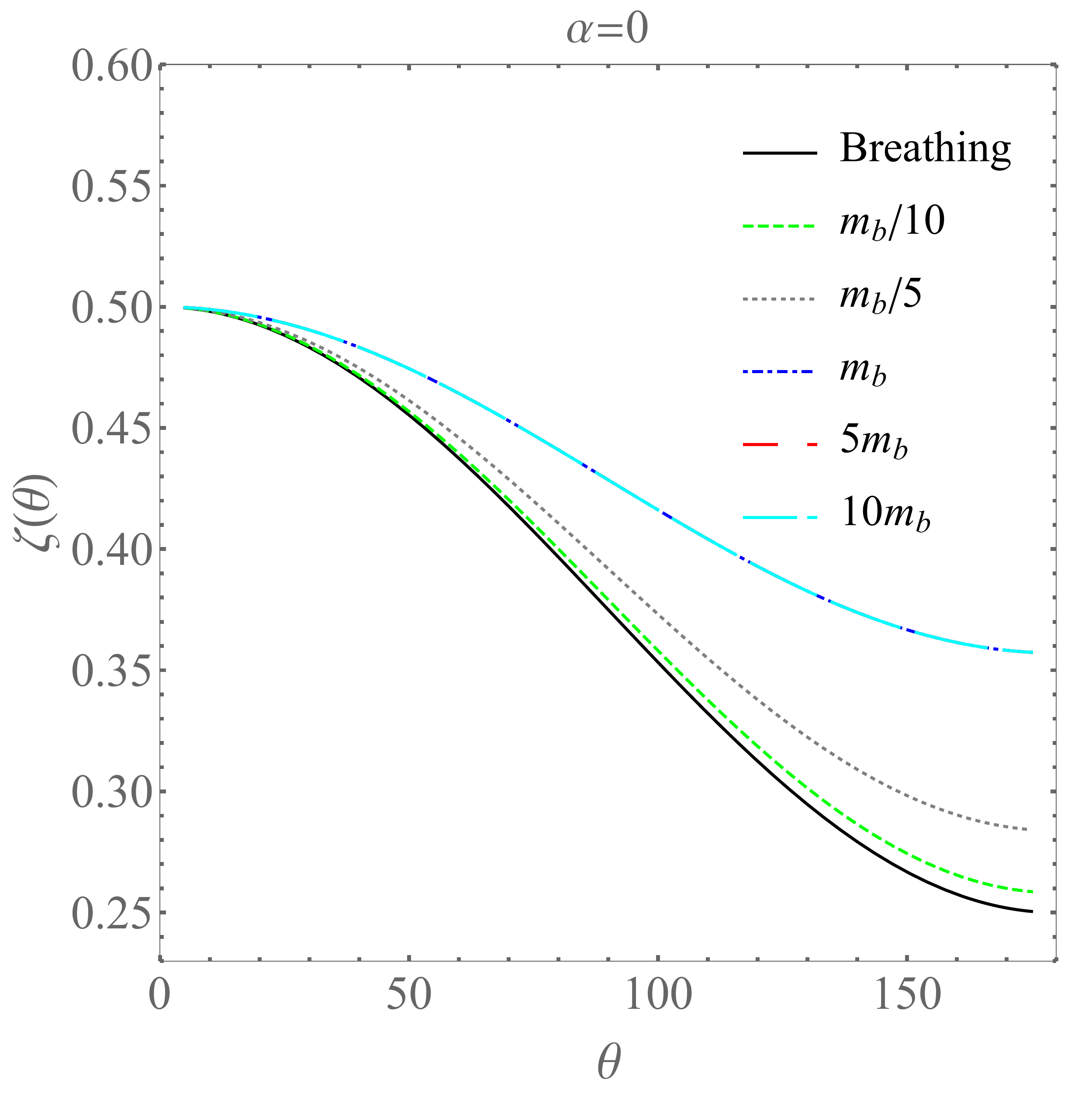}
  \caption{$\zeta(\theta)$ as a function of scalar mass $m$ at $\alpha=0$.
  The calculations were also done for a 5-year observation. Taken from Ref.~\cite{Hou:2017bqj}.}\label{fig-alpha0}
\end{figure}

A second method to detect GWs is to use pulsar timing arrays (PTAs)
\cite{Hellings:1983fr,Lee2008ptac,Lee:2010cg,Chamberlin:2011ev,Lee:2014awa,Gair:2014rwa,Gair:2015hra}.
The stochastic GW background causes the pulse time-of-arrival (TOA) residuals $\tilde{R}(t)$ from pulsars which can be measured by PTAs \cite{Hellings:1983fr}.
The TOA residuals of two pulsars (named $a$ and $b$) are correlated, which is measured by the cross-correlation function  $C(\theta)=\langle \tilde{R}_a(t)\tilde{R}_b(t)\rangle$ with $\theta$ is the angular separation between $a$ and $b$.
The brackets indicate the ensemble average over the stochastic background.
Figure~\ref{fig-normcorr} shows the normalized correlation function $\zeta(\theta)=C(\theta)/C(0)$.
The left panel shows $\zeta(\theta)$ induced by the massless field $\tilde h_{\mu\nu}$ (the solid black curve)  and the \emph{massless} scalar field $\varphi$ (the dashed blue curve).
The right panel displays $\zeta(\theta)$ induced by the mixed polarization of the transverse and longitudinal ones.
$\alpha$ is the power-law index \cite{Lee2008ptac}.
So this result provides the possibility to determine the polarization content of GWs.
In Fig.~\ref{fig-alpha0}, we calculated $\zeta(\theta)$ for the massless (labeled by Breathing) and the massive (5 different masses in units of $m_b$) cases.
One can find out that $\zeta(\theta)$ induced by $\varphi$ is quite sensitive to small masses with $m\le m_b$, but for  larger masses, $\zeta(\theta)$ barely changes.
In our approach, it is not allowed to calculate the cross-correlation function separately for the longitudinal and the transverse polarizations, because they are both excited by the same field $\varphi$ and the polarization state is a single mode.

\section{Gravitational Wave Polarizations in Einstein-\ae{}ther Theory and Generalized TeVeS Theory}\label{sec-eae-gtvs}

Finally, we briefly talk about the GW polarization contents in Einstein-\ae{}ther theory \cite{Jacobson:2004ts} and generalized TeVeS Theory \cite{Bekenstein:2004ne,Seifert:2007fr,Sagi:2010ei}.
There are more d.o.f. in these theories, and they excite more polarizations.
These two theories both contain the normalized timelike vector fields, so the local Lorentz invariance is violated.
This allows superluminal propagation.
Although all polarizations are massless, NP formalism cannot be applied neither.
The experimental constraints and the implications for the future experimental tests of these theories can be found in Ref.~\cite{Gong:2018cgj,Hou:2018djz}.

\subsection{Einstein-\ae{}ther Theory}

Einstein-\ae{}ther theory contains the metric tensor $g_{\mu\nu}$ and the \ae ther field $u^\mu$ to mediate gravity.
The action is
\begin{equation}\label{aeact}
\begin{split}
  S_{\text{EH-\ae}} =&\frac{1}{16\pi G} \int\ud^4x\sqrt{-g}[R-c_1(\nabla_\mu u_\nu)\nabla^\mu u^\nu-c_2(\nabla_\mu u^\mu)^2-c_3(\nabla_\mu u_\nu)\nabla^\nu u^\mu\\
  &+c_4(u^\rho\nabla_\rho u^\mu)u^\sigma\nabla_\sigma u_\mu+\lambda(u^\mu u_\mu+1)],
  \end{split}
\end{equation}
where $\lambda$ is a Lagrange multiplier and $G$ is the gravitational coupling constant,
the constants $c_i\,(i=1,2,3,4)$ are the coupling constants.
A special solution solves the equations of motion, i.e., $g_{\mu\nu}=\eta_{\mu\nu}$ and $u^\mu=\underline{u}^\mu=\delta^\mu_0$.
Linearizing the equations of motion ($g_{\mu\nu}=\eta_{\mu\nu}+h_{\mu\nu}$ and $u^\mu=\underline u^\mu+v^\mu$), and using the gauge-invariant variables defined in Ref.~\cite{Gong:2018cgj}, one obtains the following equations of motion
\begin{gather}
  \frac{c_{14}}{2-c_{14}}[c_{123}(1+c_2+c_{123})-2(1+c_2)^2]\ddot{\Omega}+c_{123}\nabla^2\Omega=0, \label{eomscl}\\
  c_{14}\ddot\Sigma_j-\frac{c_1-c_1^2/2+c_3^2/2}{1-c_{13}}\nabla^2\Sigma_j=0,\label{eomsig} \\
  \frac{1}{2}(c_{13}-1)\ddot h_{jk}^\mathrm{TT}+\frac{1}{2}\nabla^2h_{jk}^\mathrm{TT}=0,\label{ettjk}
\end{gather}
where $c_{13}=c_1+c_3$, $c_{14}=c_1+c_4$, and $c_{123}=c_1+c_2+c_3$.
There are five propagating d.o.f., and they propagate at three speeds.
The squared speeds  are given by
\begin{equation}
  s_g^2=\frac{1}{1-c_{13}},\quad
  s_v^2=\frac{c_1-c_1^2/2+c_3^2/2}{c_{14}(1-c_{13})},\quad
  s_s^2=\frac{c_{123}(2-c_{14})}{c_{14}(1-c_{13})(2+2c_2+c_{123})}.\label{sclspd}
\end{equation}
These speeds are generally different from each other and the speed of light.
In fact, the lack of the gravitational Cherenkov radiation requires them to be superluminal \cite{Elliott:2005va}.

The polarization content can be obtained in terms of the gauge-invariant variables
\cite{Flanagan:2005yc},
\begin{equation}\label{eq-rtitkginv}
  R_{tjtk}=-\frac{1}{2}\ddot h_{jk}^\mathrm{TT}+\dot\Xi_{(j,k)}+\Phi_{,jk}-\frac{1}{2}\ddot\Theta\delta_{jk}.
\end{equation}
Again, assume the GW propagates in the $+z$ direction with the following wave vectors
\begin{equation}
  k_s^\mu=\omega_s(1,0,0,1/s_s),\quad
   k_v^\mu=\omega_v(1,0,0,1/s_v),\quad
    k_g^\mu=\omega_g(1,0,0,1/s_g),\label{aewv-g}
\end{equation}
for the scalar, vector and tensor GWs, respectively.
One finds out that there are five polarization states:
the plus polarization is represented by $\hat P_+=-R_{txtx}+R_{tyty}=\ddot h_+$, and the cross polarization is $\hat P_\times=R_{txty}=-\ddot h_\times$;
the vector-$x$ polarization is donated by $\hat P_{xz}=R_{txtz}\propto\partial_3\dot\Sigma_1$,
and the vector-$y$ polarization is $\hat P_{yz}=R_{txty}\propto\partial_3\dot\Sigma_2$;
the transverse breathing polarization is specified by $\hat P_b=R_{txtx}+R_{tyty}\propto\dddot\Omega$,
and the longitudinal polarization is $\hat P_l=R_{tztz}\propto\dddot\Omega$.
Note that both the transverse breathing and the longitudinal modes are excited by the same scalar d.o.f. $\Omega$, so $\Omega$ excites a mixed state of $\hat P_b$ and $\hat P_l$, as in the case of Horndeski theory \cite{Hou:2017bqj,Gong:2017bru}.

Although the five polarizations are null, the NP formalism cannot be applied, as they propagate at speeds other than 1.
Indeed, the calculation showed that none of the NP variables vanish in general.

\subsection{Generalized TeVeS Theory}

Tensor-Vector-Scalar (TeVeS) theory was the relativistic realization of Milgrom's modified Newtonian dynamics (MOND) \cite{Bekenstein:2004ne,Milgrom:1983ca,Milgrom:1983pn,Milgrom:1983zz}.
It has an additional scalar field $\sigma$ to mediate gravity.
The action for the vector field $u^\mu$ is of the Maxwellian form.
Later, it was generalized and replaced by the action for the \ae ther field to solve some of the problems that TeVeS theory suffers \cite{Seifert:2007fr}.
The new theory is simply called the generalized TeVeS theory, whose action includes Eq.~\eqref{aeact} and the one for the scalar field,
\begin{equation}\label{eq-act-sig}
  S_\sigma=-\frac{8\pi}{\jmath^2\ell^2G}\int\ud^4x\sqrt{-g}\mathcal F(\jmath\ell^2 j^{\mu\nu}\sigma_{,\mu}\sigma_{,\nu}),
\end{equation}
where $j^{\mu\nu}=g^{\mu\nu}-u^\mu u^\nu$,  $\jmath>0$ is dimensionless and $\ell$ is a constant with the dimension of length.
The dimensionless function $\mathcal F$ is chosen to produce the relativistic MOND phenomena.

We use the similar method to obtain the polarization content for this theory as for Einstein-\ae{}ther theory.
Since there is one more d.o.f., there is one more polarization state: a mix polarization of the longitudinal and transverse breathing polarizations excited by the new d.o.f. $\sigma$.
This polarization state is also massless and propagates at a different speed from 1.
So the NP formalism cannot be applied to this theory either.

\section{Conclusion}\label{sec-con}

In this talk, we discussed the polarization contents in several alternative theories of gravity: $f(R)$ gravity, Horndeski theory, Einstein-\ae{}ther theory and generalized TeVeS theory.
Each theory predicts at least one extra polarization states due to the extra d.o.f. contained in it.
In the case of the local Lorentz invariant theories, such as $f(R)$ gravity and Horndeski theory, the massive scalar field excites a mix of the longitudinal and the transverse breathing polarizations;
the massless scalar field excites only the transverse breathing one.
In the case of the local Lorentz violating theories, such as Einstein-\ae{}ther theory and generalized TeVeS theory, each of the scalar d.o.f. is massless, but it propagates at speeds different from 1, so it also excites a mix of the longitudinal and the transverse breathing polarizations.
Einstein-\ae{}ther theory and generalized TeVeS theory also have vector polarizations due to the presence of the vector fields.
$E(2)$ classification was designed to categorize the polarizations for the null and Lorentz invariant theories, so it cannot be applied to these theories.
The experimental tests of the extra polarizations were also discussed.
The analysis showed that the interferometers are not sensitive to the longitudinal polarization which might be detected using PTAs.

\begin{acknowledgments}
This research was supported in part by the Major Program of the National Natural Science Foundation of China under Grant No. 11690021 and the National Natural Science Foundation of China under Grant No. 11475065.
We also thank Cosimo Bambi for the organization of the conference \emph{International Conference on Quantum Gravity} that took place in Shenzhen, China, 26-28 March, 2018. This paper is based on a talk given at the mentioned conference.
\end{acknowledgments}

%\bibliographystyle{unsrt}
%\bibliographystyle{apsrev4-1}
%\bibliography{../../../References/references}

\end{document}